\documentclass[aps,superscriptaddress,twocolumn,twoside,floatfix,prl]{revtex4-1}
\usepackage{graphicx,amsmath,amssymb,amsfonts,color,array, tikz}
\usepackage{times,comment,enumitem}
\usepackage[ocgcolorlinks,colorlinks=true,linkcolor=blue,citecolor=red,linktocpage=true,breaklinks=true]{hyperref}
\usepackage[english]{babel}
\newcommand{\ket}[1]{|#1\rangle}
\newcommand{\bra}[1]{\langle#1|}

\newcommand{\bM}{\mathbb{M}}

\DeclareMathOperator{\tr}{tr}

\begin{document}

\title{The complexity of compatible measurements}

\author{Paul Skrzypczyk}
\affiliation{H. H. Wills Physics Laboratory, University of Bristol, Tyndall Avenue, Bristol, BS8 1TL, UK.}

\author{Matty J. Hoban}
\affiliation{Department of Computing, Goldsmiths, University of London, New Cross, London SE14 6NW, United Kingdom}

\author{Ana Bel\'en Sainz}
\affiliation{International Centre for Theory of Quantum Technologies, University of Gda\'nsk, 80-308 Gda\'nsk, Poland}
\affiliation{Perimeter Institute for Theoretical Physics, 31 Caroline St. N, Waterloo, Ontario, Canada, N2L 2Y5.}

\author{Noah Linden}
\affiliation{School of Mathematics, University of Bristol, University Walk, Bristol, BS8 1TW, UK}

\begin{abstract}
Measurement incompatibility is one of the basic aspects of quantum theory. Here we study the structure of the set of compatible -- i.e. jointly measurable -- measurements. We are interested in whether or not there exist compatible measurements whose parent is maximally complex -- requiring a number of outcomes exponential in the number of measurements, and related questions. Although we show this to be the case in a number of simple scenarios, we show that generically it cannot happen, by proving an upper bound on the number of outcomes of a parent measurement that is linear in the number of compatible measurements. We discuss why this doesn't trivialise the problem of finding parent measurements, but rather shows that a trade-off between memory and time can be achieved. Finally, we also investigate the complexity of extremal compatible measurements in regimes where our bound is not tight, and uncover rich structure.
 
\end{abstract}

\maketitle

\section{Introduction}
One of the first lessons one typically learns about quantum mechanics is that in general observables do not commute, and that this leads, for example, to the famous uncertainty principle \cite{heisenberg1927}. The non-commutativity of observables shows that not all properties of a system can be measured simultaneously in quantum theory. Only when observables commute does a measurement in their common eigenbasis allow for the outcome of both measurement to be obtained simultaneously. The lack of commutativity is an indication that, in general, measurements are \emph{incompatible}. 

From a modern perspective, not all measurements are projective, and commutativity no longer fully captures the notion of incompatibility. In particular, any set of positive operators $\bM = \{M_a\}_a$ that sum to the identity, $\sum_a M_a = \openone$, constitutes a valid measurement, known as a Positive-Operator-Valued-Measure (POVM) measurement. Although for POVMs there are multiple notions of compatibility, the most important is the notion of \emph{joint measurability} \cite{busch1996,heinosaari2016}. This concerns whether there exists a single `parent' measurement which can be measured in place of the individual measurements, and which can be used to determine all their outcomes. 

Parent measurements are in general more complex than their children. In particular, they will generally have many more outcomes than their children. Our interest here is in how complex parent measurements need to be in general. While it is known that parents need to have at most a number of outcomes that is exponential in the number of measurements, it is not known to what extent this bound is tight. The exponential number of outcomes makes determining whether a set of measurements is compatible, and finding a parent, difficult. In this paper we instigate the study of this and related questions. These questions probe the structure of the set of compatible quantum measurements, a topic which has to date received very little attention \cite{haapasalo2014,guerini2018}. We give examples of measurements whose parents are maximally complex, however our main result is to find a new upper bound on the complexity of parent measurements. Our bound, which is linear in the number of measurements, is generically significantly tighter than the exponential bound. We show nevertheless why the problem of determining whether a set of measurements is compatible, and finding the corresponding parent, remains difficult even in light of our new bound. 

\section{Joint-measurability}
Consider a pair of POVM measurements $\mathbb{M} = \{M_a\}_a$ and $\mathbb{N} = \{N_b\}_b$ with $m$ and $n$ outcomes respectively. This pair of measurements is said to be jointly measurable if there exists a measurement $\mathbb{K} = \{K_\lambda\}_\lambda$ with $k$ outcomes, $\lambda \in [k] \equiv \{1,\ldots,k\}$, and a pair of partitions of $[k]$ (non-empty, disjoint subsets), $(P^\mathbb{M}_a)_a$ and $(P^\mathbb{N}_b)_b$  such that
\begin{align}\label{e:simple parent}
	M_a &= \sum_{\lambda \in P_a^\mathbb{M}} K_\lambda,& N_b &= \sum_{\lambda \in P_b^\mathbb{N}} K_\lambda. 
\end{align}
That is, instead of measuring $\mathbb{M}$ or $\mathbb{N}$ individually, the measurement $\mathbb{K}$ can be  performed, and the outcomes $a$ and $b$ generated purely classically as a function of the the outcome $\lambda$.

In \eqref{e:simple parent} it can be seen that each POVM element of $\mathbb{M}$ and $\mathbb{N}$ are sums of POVM elements of $\mathbb{K}$. For each value of $\lambda$ we can therefore specify which outcome $a$ and $b$ occurs for $\mathbb{M}$ and $\mathbb{N}$ respectively. This shows that there is a \emph{canonical} form of parent measurement $\mathbb{C} = \{C_{ab}\}_{a,b}$ which has $mn$ outcomes, such that $C_{ab}$ is associated to the outcome $a$ for $\mathbb{M}$ and $b$ for $\mathbb{N}$, i.e., such that
\begin{align}\label{e:canonical pair}
M_a &= \sum_b C_{ab},& N_b &= \sum_a C_{ab}\,. 
\end{align}
This highlights that we never need to consider parent measurements with more than $mn$ outcomes, since if two elements had the same label $ab$ when put into canonical form, they are simply added together.

More generally, a collection of $m$ measurements $\{\mathbb{M}_x\}_x$, indexed by $x$, such that $\mathbb{M}_x = \{M_{a_x|x}\}_a$, is said to be jointly measurable if there exists a parent $\mathbb{C} = \{C_\mathbf{a}\}_\mathbf{a}$, with outcomes indexed by the tuple $\mathbf{a} = (a_1,a_2,\ldots,a_m)$, such that
\begin{equation}\label{e:canonical}
	M_{a_x|x} = \sum_{\mathbf{a} / a_x} C_\mathbf{a},
\end{equation}
where $\mathbf{a} / a_x = (a_1, \ldots, a_{x-1},a_{x+1},\ldots,a_m)$. That is, the children are now arise by marginalising over all but one of outcomes of the parent. 

One notable implication of the canonical form is that the problem of deciding whether a set of measurements is compatible or not can be seen to be an instance of semidefinite programming (SDP) \cite{boyd2004}, since the conditions \eqref{e:canonical} that ensure $\mathbb{C}$ is a parent measurement are linear equality constraints, which need to be imposed in addition to the normalisation and positivity conditions that ensure $\mathbb{C}$ is a valid POVM.

As we will discuss in the next section, the canonical form also has implications for the complexity of joint measurability. 

\section{The complexity of compatibility}
An important first observation is that there is an upper bound on the number of outcomes that a parent measurement needs to have, if it exists. In particular, every parent can be written in canonical form, with outcomes labelled by the tuple $\mathbf{a}$.  If each of the measurements $\mathbb{M}_x$ has $o$ outcomes, then the there are only $o^m$ tuples, and hence only $o^m$ outcomes. 

At first sight, this might seem to appear that parents are inherently more complex than their children, having a number of outcomes that is exponential in the number of measurements. 

There are two reasons why this conclusion is premature. First, if $C_\mathbf{a}$ vanishes for some values of $\mathbf{a}$, then these outcomes never occur -- independent of the state being measured. Thus, the number of outcomes of the parent are really only the number of non-vanishing POVM elements, which could be substantially smaller than $o^m$. 

Second, a given set of measurements does not uniquely determine a parent measurement. In general there are infinitely many parent measurements for a given set of compatible measurements. This is analogous to the fact that there are in general infinitely many joint probability distributions consistent with a set of marginal probabilities. Thus, even if for a given parent none of the POVM elements vanish, this does not rule out the possibility of there being another parent, such that some of its POVM elements vanish. 

In order to understand how complex parent measurements are, we thus need to study the simplest possible parent for a given set of compatible measurements. 

More generally, this leads to a number of questions concerning the structure of the compatible measurements as a whole, which arise from the idea of studying the complexity of parent measurements: (i) Do there exist sets of compatible measurements such that  the simplest parent measurement necessarily has all elements non-vanishing? (ii) If this is not the case, then what is the exact dependence on the number of measurements $m$ in the worst case? i.e., is the complexity exponential in $m$, or might it be only polynomial? (iii) Which sets of measurements have the simplest parents, and how simple are they (excluding trivial cases, such as when the measurements $\mathbb{M}_x$ are equal, or a coarse graining of each other)? (iv) For typical sets of measurements -- i.e., measurements chosen at random with respect to a natural measure, are parents typically complex or simple?

In the remainder of this paper we will present our initial findings concerning these questions. We believe that these questions, along with the many related ones that they lead to, collectively constitute an interesting line of investigation which has hitherto been unaddressed. 

\section{From parent measurements to compatible children}
It will be instructive to start our investigation on the complexity of compatibility by focusing on parent measurements of the form \eqref{e:simple parent}, and their resulting children.

Consider as a simple concrete example the parent measurement $\mathbb{K}$ with POVM elements 
\begin{equation}
K_\lambda = \frac{1}{3}\left(\openone + \hat{\mathbf{n}}_\lambda\cdot \boldsymbol{\sigma}\right),
\end{equation}
for $\lambda = 0,1,2$, where $\boldsymbol{\sigma} = (X,Y,Z)$ is the vector of Pauli operators, and $\hat{\mathbf{n}}_\lambda = (\cos (2\lambda\pi/3), \sin (2\lambda\pi/3), 0)$. Such a measurement -- referred to as the `trine' -- has outcomes corresponding to 3 vertices of an equilateral triangle in the $xy$-plane of the Bloch sphere. The trine measurement leads straightforwardly to a set of 3 compatible measurements, $\mathbb{L}$, $\mathbb{M}$ and $\mathbb{N}$, via 
\begin{align}\label{e:trine}
L_0 &= K_0,& M_0 &= K_1,& N_0 &= K_2,\\
L_1 &= K_1 + K_2,& M_1 &= K_0 + K_2,& N_1 &= K_0 + K_1\,,\nonumber
\end{align}
which are just the three distinct partitions of the three-outcome parent into two-outcome children. This constitutes a set of 3 two-outcome measurements that has a simple parent with only 3 outcomes. The complexity of such a parent then is much simpler than that of the canonical parent that gives the exponential upper bound $2^3 = 8$ on the number of outcomes. Notice also that this moreover provides an example of a set of 3 measurements which are a compatible set, despite not commuting.

This example can be generalised, and it is always possible to form children from a parent by considering all of the distinct partitions of the parents outcomes. The total number of partitions of a set of size $o$ is given by the Bell number $B_o$ \cite{bell1934}. Discounting the two trivial partitions (the parent itself, and the trivial measurement with a single outcome), a parent with $o$ outcomes is always a parent for a set of measurements of size $B_o - 2$. 
The Bell numbers grow rapidly, which has crucial consequences for the complexity of compatible measurements. Take for instance the case of a parent with 6 outcomes. Here $B_6 = 203$, hence every six-outcome measurement leads to set of 201 measurements that are compatible. Therefore, these 201 compatible measurements may arise from a six-outcome parent, which is overwhelmingly simpler than the canonical parent (upper bound), whose number of elements is exponential in the number of measurements. 

\section{Maximally complex sets of compatible measurements}
We now turn our attention to the other direction,  and study whether there exist sets of compatible measurements with the opposite behaviour: to be maximally complex, i.e., such that the only possible parent POVM for the set necessarily contains the maximal number of $o^m$ outcomes.

In the simplest case, of two measurements with two outcomes, it is straightforward to show that maximally complex sets of measurements exist. Consider noisy Pauli measurements, 
\begin{align}\label{e:noisy pauli}
M_{a} &= \tfrac{1}{2}(\openone + \eta(-1)^a X),& N_b &= \tfrac{1}{2}(\openone + \eta(-1)^b Z)
\end{align}
where $X$ and $Z$ are the Pauli operators. Such measurements are compatible for $\eta \leq 1/\sqrt{2}$ \cite{}. A parent measurement for $\eta = 1/\sqrt{2}$ is given by
\begin{equation}
C_{ab} = \frac{1}{4}\left( \openone + \frac{(-1)^{a}X + (-1)^{b}Z}{\sqrt{2}}\right).
\end{equation}
This parent has 4 non-zero elements. Let us assume that there exists a parent with only 3 non-zero elements. Let us assume therefore that $C_{00} = 0$ (all other cases follow an identical logic). In order to reproduce the correct POVM elements of $M_a$ and $N_b$, we must have $M_{0} = C_{00} + C_{01} = C_{01}$ and $N_{0} = C_{00} + C_{10} = C_{10}$ and in order to be normalised we must have $\openone = C_{00} + C_{01} + C_{10} + C_{11} = M_{0} + N_{0}+ C_{11}$. This implies that 
\begin{equation}
C_{11} = \openone - M_{0} - N_{0} \nonumber = -\eta (X + Z).
\end{equation}
This operator is however not positive semidefinite for any $\eta \neq 0$, with negative eigenvalue -$\sqrt{2}\eta$. Thus, there is no parent POVM with 3 elements for the set of measurements in Eq.~\eqref{e:noisy pauli}.

The above example was particularly simple because there is a unique solution for the parent. In general, a parent has $o^m$ elements, and there are $(o-1)m + 1$ linearly independent constraints 
that these elements should satisfy to reproduce the given measurements: a constraint from all but one of the POVM elements of each of the measurements $\mathbb{M}_x$, in addition to the normalisation constraint. Thus, the only situation where a unique parent POVM exists is when $o^m - (o-1)m -1$ elements of the parent vanish, which is to say that it has $(o-1)m + 1$ outcomes. The above situation where $o = m = 2$ is the only situation where $(o-1)m+1 = o^m - 1$, i.e. where a unique solution exists under the assumption that the parent is not maximally complex. 

Moving on to the next simplest case, that of $m = 3$ measurements with $o = 2$ outcomes, we were able to find an example for $d=3$, i.e., for qutrit measurements. In particular, consider the following measurements
\begin{align} \label{e:3d example}
L_{a} &= \tfrac{1}{2}\big(\openone_3 + (-1)^a(\tfrac{3\sqrt{2}}{8}X_{01}+ \tfrac{1}{2}X_{02})\big) \,, \nonumber \\
M_{b} &=  \tfrac{1}{2}\big(\openone_3 + (-1)^b(\tfrac{3\sqrt{2}}{8}Z_{01} + \tfrac{1}{2}X_{12})\big) \,,  \\
N_{c} &= \tfrac{1}{2}\big(\openone_3 + (-1)^c\tfrac{1}{2}(Z_{02}+Z_{12})\big) \,, \nonumber 
\end{align}
where $X_{ij} = \ket{i}\bra{j} + \ket{j}\bra{i}$ and $Z_{ij} = \ket{i}\bra{i} - \ket{j}\bra{j}$ are Pauli operators acting on the subspace spanned by $\ket{i}$ and $\ket{j}$. 

These measurements can be shown to be jointly measurable, with a parent with 8 non-vanishing elements. However, it can also be shown that it is not possible to find a parent with $7$ or fewer non-vanishing elements. In particular, if $C_{abc} = 0$ is imposed, for any value of $a$,$b$ and $c$, then no parent exists for this set of measurements. The proof of this claim is not as straightforward as in the previous example, and is based on a method we develep using the duality theory of semidefinite programs. 
This method, whose full details are provided in the appendix, allows us to find a  ``witness'', which guarantees that there is no parent measurement for the children if any POVM element of the parent vanishes. Full details are provided in the appendix. This shows that it is possible to certify conclusively that parents with given elements vanishing do not exist. 

Finally, by a heuristic search method we were able to numerically find examples in a number of other cases: for two measurements ($m=2$) in dimension $d$ with $o = d$ outcomes, requiring a parent with $d^2$ elements in each case. Details are provided in an accompanying online notebook \cite{notebook}.

In the next section, however, we will show that the maximal complexity of a set of measurements in fact does not grow exponentially, but actually grows much slower in the number of measurements $m$.

\section{Bounding the complexity of compatibility}
In this section we will outline a proof that the number of outcomes of a parent measurement can always be bounded from above, by considering the geometry of the set of compatible measurements. In the appendix we provide a full proof which demonstrates that for all sets of $m$ measurements $\{\bM_x\}_x$, with $o$ outcomes in dimension $d$, the maximal number of outcomes is never bigger than
\begin{equation}\label{e: bound}
d^2(m(o-1) + 1) \,,
\end{equation}
i.e. that the dependence on the number of measurements is in fact linear, and not exponential.

The basic idea behind the proof is to use the geometry of the set of compatible measurements, and to realise that it has the structure of a cone, with extremal rays given by deterministic sub-normalised measurements. Caratheodory's theorem (for cones) \cite{caratheodory1907} can then be used in order to show that no more than  $d^2(m(o-1) + 1)$ extremal rays are necessary to form an arbitrary point inside the cone (a set of compatible measurements), which, translated back, means that a parent never needs more than this many outcomes.

Note that this bound can be larger than $o^m$, which in particular happens when $m =2$ and $o = d$ (as in the previous examples), hence why maximally complex sets can be found in certain cases. However, in general this shows that as the number of measurements grows, the complexity does not increase exponentially, as would be naively thought. 

\section{Memory versus time trade-off}
A major barrier to determining whether a set of measurements is compatible or not is the exponential increase in the size of the SDP optimisation problem that has to be solved as the number of measurements increases. Since canonical parents have exponentially many outcomes in the number of measurements, it quickly becomes impractical, due to lack of memory, to determine whether a set of measurements is compatible or not. 

The bound \eqref{e: bound} however says that this exponential overhead is not required; in fact, a much smaller parent can be found. Potentially the bound then implies that a more efficient algorithm could be found in order to determine whether a set of measurements is compatible or not.

Crucially however, although this bound implies the existence of a much smaller parent, a new problem arises: determine which elements of the canonical parent are non-vanishing. The bound implies that it is sufficient to check the $\binom{o^n}{d^2(m(o-1) + 1)}$ parents with $d^2(m(o-1) + 1)$ non-zero elements in order to determine whether the measurements are incompatible, or not. 

Thus the need for a large memory can be overcome, at the expense of needing to carry out a significant number of calculations. In situations where calculations can easily be run in parallel, this may provide a practical way to attack problem instances that were previously infeasible due to memory requirements.

\section{Complexity of typical measurements}
A further interesting question is to understand the behaviour of typical sets of compatible measurements. In scenarios where the bound \eqref{e: bound} is tight, we see immediately that typical sets of compatible measurements will be maximally complex. This follows from the geometry of the problem -- since typical measurements will live in the interior of the set (for any reasonable measure), they will need to be conic combinations of $d^2(m(o-1) + 1)$ points, and not fewer. 

\begin{figure}[t!]
\centering 
\includegraphics[width=\columnwidth]{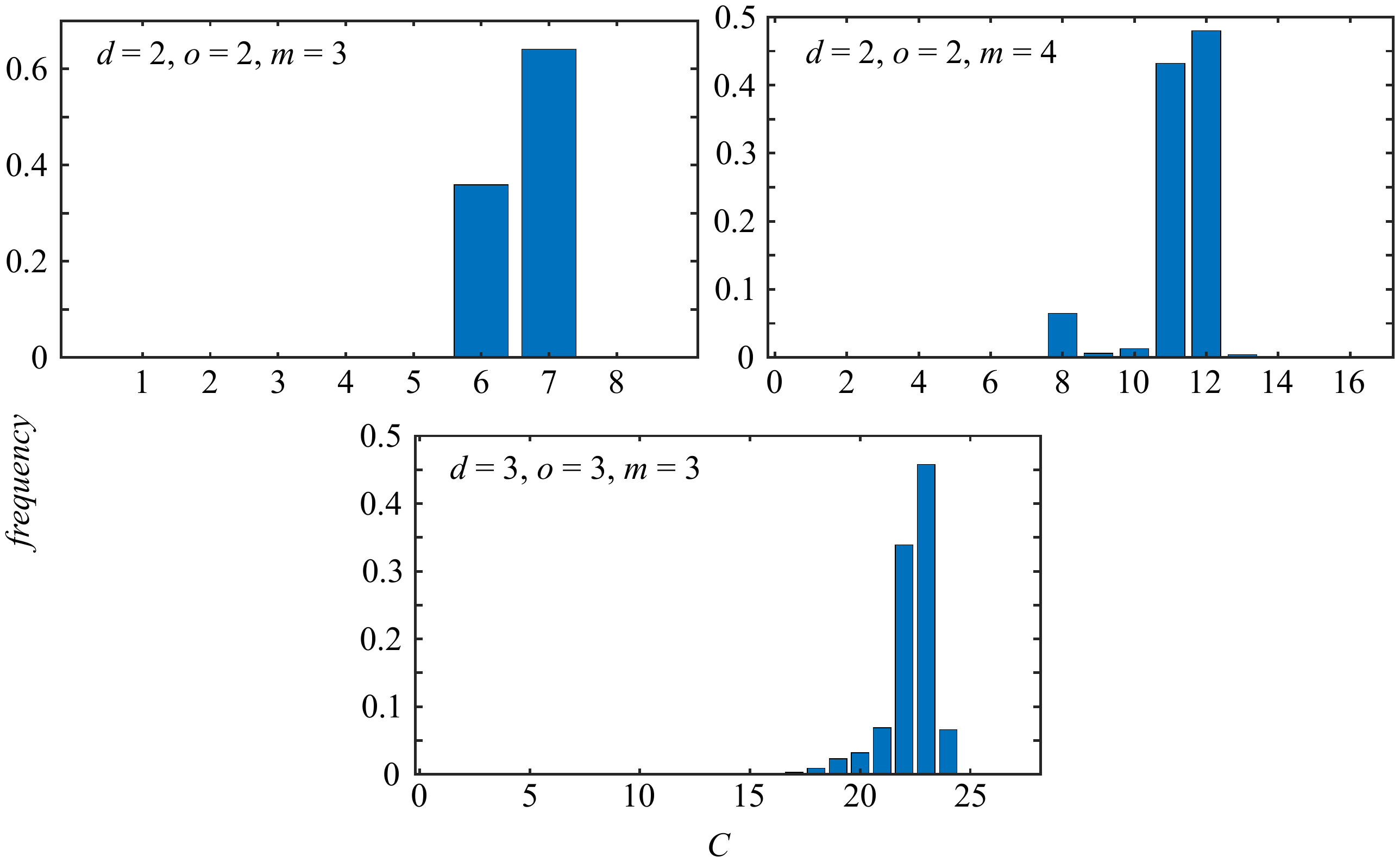}
\caption{Histograms showing the distribution of complexity for measurements sampled from the boundary of the jointly measurable set using the induced measure (as described in the appendix). In each instance, 1000 points were sampled, and we give here the relative frequencies of the complexity of the measurements, computed numerically. Interestingly, in all cases, no maximally complex measurements are found. }
\label{f:typical meas results}
\end{figure}

In scenarios where the bound \eqref{e: bound} is not tight, the situation is much less clear. It would be interesting to know whether most sets of compatible measurements have simple parents or not. One problem with tackling this directly is the need to generate random instances of compatible measurements, according to some measure. Here we  focus on the typical behaviour of sets of measurements which are on the boundary of the set of compatible measurements. In this case, there are natural ways to induce random measures on the boundary, based upon the Haar measure on unitary matrices. Details of this method can be found in the appendix. A summary of the results are presented in Fig.~\ref{f:typical meas results}. We find a rather complicated structure -- in all cases we find a distribution of parent sizes, indicating that the boundary has rich structure. 

\section{Complexity of EPR-steering}
The above also imply results about the complexity of Local-Hidden-State (LHS) models in the context of Einstein-Podolsky-Rosen (EPR) steering \cite{wiseman2007,cavalcanti2017,uola2019a} 

EPR steering is the nonlocal effect whereby measurements performed by Alice on half of an entangled quantum state, `steer' the states of Bob at a distance in a way which cannot be explained by a simple (essentially classical) causal model (LHS model). In particular, if Alice and Bob share a state $\rho^{AB}$, and Alice performs a measurement $\bM_x$, then upon obtaining outcome $a$ she prepares for Bob the unnormalised state $\sigma_{a|x} = \tr_A[(M_{a|x}\otimes \openone) \rho^{AB}]$. The collection of states is said to have an LHS model if $\sigma_{a|x} = \sum_\lambda p(a|x,\lambda) \sigma_\lambda$, where $p(a|x,\lambda)$ are a collection of probabilities describing Alice's measurement outcome, and $\sigma_\lambda$ are a collection of unnormalised hidden states, describing Bob's system. Similarly to the case of compatibility, one can study the complexity of LHS models, by asking for the smallest sized LHS model, i.e. the model with the smallest number of hidden states.

It was recently shown that there is a one-to-one correspondence between incompatibility and steering -- every set of measurements leads to steering if and only if it is incompatible \cite{quintino2014,uola2014}. In fact, starting from an LHS model in the steering scenario, it is always possible to obtain a parent measurement, and vice versa (see appendix for details). As such, the above results about the complexity of compatibility imply that the complexity of LHS models is identical and that LHS models never need to contain more than $d^2(m(o-1) + 1)$ hidden states. 

\section{Probabilistic Parents}
We finally discuss the possibility of a more general class of parent measurements, and the implications this may have for complexity. When introducing joint measurability, our general definition was that children are generated from parents by partitioning the outcomes of the parent. A slightly more general definition is to introduce randomness, and allow for mixing over partitions. In this case, a pair of measurements $\mathbb{M} = \{M_a\}_a$ and $\mathbb{N} = \{N_b\}_b$ are jointly measurable if there exist a parent measurement $\mathbb{K} = \{K_\lambda\}_\lambda$ and a pair of conditional probability distributions $p_\mathbb{M}(a|\lambda)$ and $p_\mathbb{N}(b|\lambda)$ such that
\begin{align}\label{e:prob parent}
	M_a &= \sum_\lambda p_\mathbb{M}(a|\lambda)K_\lambda, & N_b &= \sum_\lambda p_\mathbb{N}(b|\lambda) K_\lambda. 
\end{align}
This definition does not change whether or not a set of measurements is compatible or not with respect to our previous notion, since it is always possible to go from this form to the canonical form, as we briefly outline in the appendix. It also does not affect our upper bound \eqref{e: bound}, since this provides an explicit parent for a set of children. It does however have implications for the existence of maximally complex sets of compatible measurements, since in principle it might be possible to reduce the size of the parent by using randomness. 

We find that this is indeed the case. For the example  given in Eq.~\eqref{e:noisy pauli} we were able to find the following probabilistic parent with only 3 outcomes:
\begin{align}
K_0 &= \frac{2 M_0 - \eta \sqrt{2} N_0}{2-\eta^2}, & K_1 &= \frac{2 N_0 - \eta \sqrt{2} M_0}{2-\eta^2},
\end{align}
 and $K_2 = \openone - K_0 - K_1$. Indeed, $\{K_1, K_2, K_3\}$ is a valid POVM for $\eta \lessapprox 0.5609$, and reproduces $\mathbb{M}$ and $\mathbb{N}$ with 
\begin{align}
 M_0 &= K_0 + \tfrac{\eta}{\sqrt{2}}K_1,& N_0 &= \tfrac{\eta}{\sqrt{2}}K_0 + K_1.
\end{align} 
We do not know whether or not a probabilistic parent exists for larger values of $\eta \leq 1/\sqrt{2}$. Nevertheless, no deterministic parent with 3 outcomes exists for any $\eta$, as previously shown.  

The nature of the problem of finding probabilistic parents seems much richer than the problem of finding deterministic parents, which has the simple form of a convex optimisation problem. When introducing probabilistic parents, this convex structure is lost, and it seems much harder to decide whether a probabilistic parent of a given size exists. Furthermore, this raises an interesting question regarding the possible trade-off between the amount of randomness necessary and the size of the parent. We leave further exploration of these interesting questions for future work. 

\section{Conclusions}
In this work we have instigated the study of the complexity of a set of compatible measurements. In one direction we have shown that very large sets of compatible measurements can be formed starting from a parent, and in the other direction that the complexity of compatibility can be bounded, and scales no worse than linearly in the number of measurements $m$. We have also explored the typical behaviour of the boundary of compatible measurements in instances where this bound is not tight, and shown that in these cases the boundary appears to have a rich structure. We have finally raised the possibility of using randomness in the parent to reduce the complexity, and found examples where this is indeed possible. 

We believe our results raises many interesting questions about the structure of compatible measurements that should lead to further exciting work in this and related directions.

\section{Acknowledgements}
PS acknowledges support from a Royal Society URF (UHQT). ABS acknowledges support by the Foundation for Polish Science (IRAP project, ICTQT, contract no. 2018/MAB/5, co-financed by EU within Smart Growth Operational Programme). This research was partially supported by Perimeter Institute for Theoretical Physics. Research at Perimeter Institute is supported by the Government of Canada through the Department of Innovation, Science and Economic Development Canada and by the Province of Ontario through the Ministry of Research, Innovation and Science. MJH and ABS acknowledge the FQXi large grant “The Emergence of Agents from Causal Order”.

\bibliography{jointly-meas-arxiv}

\begin{appendix}
\section{APPENDICES}

\section{Appendix A: Method for certifying complexity}
In this appendix we provide details for how one can certify a lower bound on the complexity of a set of compatible measurements, using the duality theory of semidefinite programming. We first outline how the problem of compatibility can be written as an SDP, and demonstrate that this remains true even when fixing the size of the parent. We then show how the dual of this modified SDP leads to certificates that a set of measurements doesn't have a parent of a given size. We apply this method to the example from the main text, to show that it can be put to use to explicitly demonstrate that a given set of compatible measurements has maximal complexity. 

\subsection{Primal and dual SDP formulation of joint measurability}
As stated in the main text, the canonical form of parent POVM demonstrates that the problem of deciding whether a set of measurements $\{\mathbb{M}_x\}_x$ is compatible is a feasibility semidefinite program. In particular, it is given by the problem
\begin{align}\label{e:feasible sdp}
\text{find}&&  \{C_\mathbf{a}\}_{\mathbf{a}} \nonumber \\
\text{subject to}&& M_{a|x} &= \sum_\mathbf{a} D_\mathbf{a}(a|x) C_\mathbf{a} \quad \forall a,x,\nonumber \\
&& C_\mathbf{a} &\geq 0 \quad \forall \mathbf{a}, \nonumber \\
&& \openone &= \sum_\mathbf{a} C_\mathbf{a}.
\end{align}
where $D_\mathbf{a}(a|x) = \delta_{a,\mathrm{a}_x}$ can be thought of as a collection of deterministic conditional probability distributions, labelled by $\mathbf{a}$, such that $a = \mathrm{a}_x$ with certainty.

This problem is feasible if and only if the set of measurements is jointly measurable. In general it is easier to work an equivalent form, which is a semidefinite optimisation problem,
\begin{align}\label{e: optim sdp}
\max_{\nu,\{C_\mathbf{a}\}} &&  \nu \nonumber \\
\text{s.t. }&& M_{a|x} &= \sum_\mathbf{a} D_\mathbf{a}(a|x) C_\mathbf{a} \quad \forall a,x,\nonumber \\
&& \openone &= \sum_\mathbf{a} C_\mathbf{a},\nonumber \\
&& 0 &\geq \nu, \nonumber \\
&& C_\mathbf{a} &\geq \nu \openone \quad \forall \mathbf{a}.
\end{align}
where we have introduced a new non-positive scalar variable $\nu$ which is to be maximised. If $v = 0$ is a feasible for this problem, then each of the $C_\mathbf{a}$ are positive semidefinite and thus form a valid parent POVM for the measurements $\{\mathbb{M}_x\}_x$. If on the other hand, the maximal feasible value of $\nu$ is negative, then this indicates that the problem \eqref{e:feasible sdp} is infeasible, and the set of measurements $\{\mathbb{M}_x\}_x$ are incompatible. 

If we are interested in finding parent POVMs which are not maximal, we can use modified versions of the above SDPs to tackle the problem. In particular, if we denote by $O(\mathbb{C}) = \{\mathbf{a}\,|\, C_\mathbf{a} \neq 0\}$ the list of outcomes of $\mathbb{C}$, then the problem remains an SDP when we specify this list.
That is, a parent POVM will exist with the only non-zero outcomes a subset of $O(\mathbb{C})$ if and only if the value of the following SDP is zero
\begin{align}\label{e: optim sdp w null}
\max_{\nu,\{C_\mathbf{a}\}} &&  \nu \nonumber \\
\text{s.t. }&& M_{a|x} &= \sum_{\mathbf{a}\in O(\mathbb{N})} D_\mathbf{a}(a|x) C_\mathbf{a} \quad \forall a < (o-1),x\nonumber \\
&& C_\mathbf{a} &\geq \nu \openone \quad \forall {\mathbf{a}\in O(\mathbb{C})}, \nonumber \\
&& \openone &= \sum_{\mathbf{a}\in O(\mathbb{N})} C_\mathbf{a},\nonumber \\
&& 0 &\geq \nu,
\end{align}
where we recall that, due to normalisation, it is not necessary to impose the final constraint (when $a = o-1$) on each element of $M_{a|x}$, as it is automatically satisfied. 

The benefit of this optimisation form of the problem over its feasibility version, is that the dual formulation of \eqref{e: optim sdp w null} will allow us to find witnesses that certify that a given set of compatible measurements has complexity at least $c$, by proving that the value of \eqref{e: optim sdp w null} is strictly negative for all sets $O(\mathbb{C})$ of size $c-1$. That is, we can check all possible parents with a given number of elements, and if none of them provide a valid parent POVM, then the complexity must be larger than the number checked. 

We note that this requires the evaluation of $\binom{o^m}{c}$ semidefinite programs, a number which quickly becomes impractical. In the case of checking for maximal complexity, i.e., $c = o^m - 1$, it requires the evaluation of $o^m$ SDPs. 

The dual formulation of the SDP of Eq.~\eqref{e: optim sdp w null} is particularly useful, as we see next. Using standard techniques for obtaining the dual formulation of a convex optimisation problem, it can be shown that the dual formulation of \eqref{e: optim sdp w null} is given by
\begin{align}\label{e: dual optim}
\min_{\{\rho_{ax}\},\omega} && &\tr\sum_{a \neq (o-1),x} \rho_{ax} M_{a|x} + \tr \omega \nonumber \\
\text{s.t.} && \omega &+ \sum_{a\neq (o-1),x} D_\mathbf{a}(a|x)\rho_{ax} \geq 0 \quad \forall \mathbf{a} \in O(\mathbb{C}),\nonumber \\
&& 1 &\geq \tr \sum_{\substack{ a\neq (o-1),x\\ \mathbf{a} \in O(\mathbb{C})}} D_\mathbf{a}(a|x)\rho_{ax} + o_\mathbb{C} \tr \omega \,,
\end{align}
where $\{\rho_{ax}\}_{a,x}$ and $\omega$ are the dual variables and $o_\mathbb{C} = |O(\mathbb{C})|$. This dual optmisation problem is useful due to weak duality, which says that for any feasible set of dual variables $\{\rho_{ax}\}_{a,x}$ and $\omega$ (which don't even need to be optimal for the problem \eqref{e: dual optim}), then 
\begin{equation}
\tr\sum_{a \neq (o-1),x} \rho_{ax} M_{a|x} + \tr \omega \geq \nu^*
\end{equation}
where $\nu^*$ is the optimal value of \eqref{e: optim sdp w null}. Therefore, if we can find explicit choices of $\{\rho_{ax}\}_{a,x}$ and $\omega$ for each set $O(\mathbb{C})$ with $o_\mathbb{C}$ outcomes, such that $\tr\sum_{a \neq (o-1),x} \rho_{ax} M_{a|x} + \tr \omega < 0$, then this provides a certificate, or witness, that the complexity of $\{\mathbb{M}_x\}_x$ is at least $o_\mathbb{C} + 1$. 

\subsection{Certificate for $m=3$ measurements with $o=2$ outcomes}
We will now apply the above theory to prove that the example of $m=3$ measurements with $o=2$ outcomes provided in the main text requires a maximal parent with $8$ outcomes. 

The set of measurements defined in Eq.~\eqref{e:3d example} may be expressed as follows, using a slightly more general notation which will be useful later:
\begin{align} \label{e:3d example appendix}
M_{a_0|0} &= \tfrac{1}{2}\big(\openone_3 + (-1)^{a_0}(\tfrac{3\sqrt{2}}{8}X_{01}+ \tfrac{1}{2}X_{02})\big) \,, \nonumber \\
M_{a_1|1} &=  \tfrac{1}{2}\big(\openone_3 + (-1)^{a_1}(\tfrac{3\sqrt{2}}{8}Z_{01} + \tfrac{1}{2}X_{12})\big) \,, \\
M_{a_2|2} &= \tfrac{1}{2}\big(\openone_3 + (-1)^{a_2}\tfrac{1}{2}(Z_{02}+Z_{12})\big) \,, \nonumber 
\end{align}
where $X_{ij} = \ket{i}\bra{j} + \ket{j}\bra{i}$ and $Z_{ij} = \ket{i}\bra{i} - \ket{j}\bra{j}$ are Pauli operators acting on the subspace spanned by $\ket{i}$ and $\ket{j}$. These measurements are jointly measurable, and a parent with 8 elements is given by $C_{a_0a_1a_2} = \frac{3}{8}\ket{\phi_{a_0a_1a_2}}\bra{\phi_{a_0a_1a_2}}$, where
\begin{align}\label{e:3d example parent}
\ket{\phi_{a_0a_10}} &= (-1)^{a_0}\cos \tfrac{\pi}{8}\ket{a_1} + \sin \tfrac{\pi}{8}\ket{a_1 \oplus 1} \,, \nonumber \\
\ket{\phi_{a_0a_1 1}} &= \frac{\ket{0} + (-1)^{(a_0 + a_1)}\ket{1} + 2(-1)^{a_0}\ket{2}}{\sqrt{6}} \,.
\end{align}
Here we will show that it is not possible to find a parent with 7 or fewer non-vanishing elements. 

Let us assume that there is a parent with 7 elements, and denote by $\mathbf{a}^*$ the string of outcomes corresponding to the POVM element of the parent which is assumed to vanish, i.e., $C_{\mathbf{a}^*} = 0$. Consider the follow sets of dual variables
\begin{align}
\rho_{0x}^{\mathbf{a}^*} &= \frac{1}{5}(-1)^{\mathrm{a}_x^*}\ket{\psi_{\mathbf{a}^*}}\bra{\psi_{\mathbf{a}^*}} \,, \\
\omega^{\mathbf{a}^*} &= \frac{1}{5}\left((\bar{\mathrm{a}}_0^*+\bar{\mathrm{a}}_1^*)\bar{\mathrm{a}}_2^*+(\bar{\mathrm{a}}_0^*+\bar{\mathrm{a}}_1^*-1)\bar{\mathrm{a}}_2^*\right)\ket{\psi_{\mathbf{a}^*}}\bra{\psi_{\mathbf{a}^*}} \,, 
\end{align}
where $\bar{\mathrm{a}}_x^* \equiv \mathrm{a}_x^* \oplus 1$,
\begin{multline}
\ket{\psi_{\mathbf{a}^*}} = (-1)^{(\mathrm{a}_0^*+\mathrm{a}_1^*)}\alpha_{\mathrm{a}_2^*}\ket{\mathrm{a}_1^*} + \beta_{\mathrm{a}_2^*}\ket{\mathrm{a}_1^*\oplus 1} \\ + (-1)^{\mathrm{a}_0^*\mathrm{a}_1^*}\gamma_{\mathrm{a}_2^*}\ket{2} \,,
\end{multline}
and $(\alpha_{\mathrm{a}_2^*},\beta_{\mathrm{a}_2^*},\gamma_{\mathrm{a}_2^*})$ are amplitudes which will be specified shortly. Direct substitution shows that for any choice of $(\alpha_{\mathrm{a}_2^*},\beta_{\mathrm{a}_2^*},\gamma_{\mathrm{a}_2^*})$, for all $\mathbf{a}^*$,
\begin{align}
\omega^{\mathbf{a}^*} &+ \sum_{x} D_\mathbf{a}(0|x)\rho_{0x}^{\mathbf{a}^*} \geq 0 \quad \forall \mathbf{a} \neq \mathbf{a}^*,\nonumber \\
1 &\geq \tr \sum_{x, \mathbf{a} \neq \mathbf{a}^*} D_\mathbf{a}(0|x)\rho_{0x}^{\mathbf{a}^*} + 7 \tr \omega^{\mathbf{a}^*},
\end{align}
which shows that $\rho_{0x}^{\mathbf{a}^*}$ and $\omega^{\mathbf{a}^*}$ are feasible solutions for the dual problem \eqref{e: dual optim}. 

Choosing 
\begin{equation}
(\alpha_{\mathrm{a}_2^*},\beta_{\mathrm{a}_2^*},\gamma_{\mathrm{a}_2^*}) =
\begin{cases} (1,0,0)& \text{if } \mathrm{a}_2^* = 0 \\
(\tfrac{\sqrt{2}}{4},\tfrac{\sqrt{2}}{4}, \tfrac{\sqrt{3}}{2}) & \text{if } \mathrm{a}_2^* = 1
\end{cases}
\end{equation}
leads to
\begin{equation}
\tr\sum_{x} \rho_{0x}^{\mathbf{a}^*} M_{a|x} + \tr \omega^{\mathbf{a}^*} 
	= \begin{cases} \frac{4-3\sqrt{2}}{80} & \text{if } \mathrm{a}_2^* = 0 \\
					\tfrac{12-8\sqrt{6} -(-1)^{\mathrm{a}_1^*}3\sqrt{2}}{320}& \text{if } \mathrm{a}_2^* = 1
	\end{cases}
\end{equation}
which is negative in all cases. This provides the required proof that it is impossible to find a parent for the set of measurements \eqref{e:3d example appendix} with 7 or fewer elements. In particular, we have demonstrated a feasible solution to the dual problem that obtains a negative value, which implies that the solution of the primal problem is negative (and hence no parent exists). Our construction works for all 8 choices of parents with (at least) 1 element vanishing. 

\section{Appendix B: A bound on the complexity of compatibility}
In this section we provide details for how the bound \eqref{e: bound} from the main text is derived, which provides a linear upper bound on the complexity of any set of compatible measurements. The proof is geometrical in nature, and relies on Caratheodory's theorem for convex cones. 

Consider a set of $m$ compatible measurements $\{\mathbb{M}_x\}_x$ acting on $\mathbb{C}^d$, each with $o$ outcomes. Each POVM element is specified by $d^2$ real parameters. Once $o-1$ POVM elements of a given measurement are specified, then the remaining element is fixed, due to the normalisation condition $\sum_a M_{a|x} = \openone$ for all $x$. Thus, in total $d^2(m(o-1))$ real parameters are required to uniquely specify a set of measurements. In what follows, we will also redundantly keep track of the right-hand side of the normalisation condition, since we will need to relax it. That is, we will also keep track of the $d^2$ parameters necessary to specify $\openone$. As such, we can represent a set of measurements by a point in a real vector space $\mathbb{R}^D$ where $D = d^2(m(o-1) + 1)$. 

Consider now a set of compatible measurements, such that
\begin{equation}
	M_{a|x} = \sum_\mathbf{a} D_\mathbf{a} (a|x) C_\mathbf{a} \,.
\end{equation}
For each $\mathbf{a}$, we can also represent $D_\mathbf{a}(a|x) C_\mathbf{a}$ by a point in the same space $\mathbb{R}^D$. Note that for such points $\sum_a D_\mathbf{a}(a|x) C_\mathbf{a} = C_\mathbf{a}$, thus these do not represent measurements, but rather sub-measurements, since $C_\mathbf{a} \leq \openone$. This was the reason for including the extra $d^2$ parameters in the above, since this allows us to consider the space of all sub-measurements. 

Geometrically, we see that the point in $\mathbb{R}^D$ corresponding to the collection of measurements $\{\mathbb{M}_x\}_x$ is a positive combination of at most $o^n$ points, one corresponding to each $D_\mathbf{a}(a|x) C_\mathbf{a}$ (since some of the $C_\mathbf{a}$ may vanish, such points correspond to the origin in $\mathbb{R}^D$, and all these points map degenerately there).

Assume now that a parent has been found with $k > D$ outcomes, i.e. such that $k$ of the $C_\mathbf{a}$ do not vanish. We thus have $k$ points in $D$ dimensions, and hence the points cannot all be linearly independent. Therefore, there must exist numbers $\lambda_\mathbf{a}$ such that
\begin{equation}
\sum_\mathbf{a} \lambda_\mathbf{a} D_\mathbf{a}(a|x) C_\mathbf{a} = 0.
\end{equation}
We take $\lambda_\mathbf{a} = 0$ for any $\mathbf{a}$ such that $C_\mathbf{a} = 0$, such that the summation indeed runs over all values of $\mathbf{a}$, and not only the $k$ non-vanishing elements. 

Therefore, we see that the following equation holds for all $\gamma$,
\begin{equation}
M_{a|x} = \sum_\mathbf{a} (1-\gamma \lambda_\mathbf{a})D_\mathbf{a} (a|x) C_\mathbf{a} \,.
\end{equation}
This implies the existence of a new potential parent measurement $\mathbb{C}' = \{C'_\mathbf{a}\}_\mathbf{a}$ with
\begin{equation}
C'_\mathbf{a} = (1-\gamma \lambda_\mathbf{a})C_\mathbf{a}
\end{equation}
This will only be a valid measurement if $C'_\mathbf{a}\geq 0$ for all $\mathbf{a}$. Therefore, we take
\begin{equation}
\gamma = \frac{1}{\max_\mathbf{a} \lambda_\mathbf{a}}
\end{equation}
such that for all $\mathbf{a}$ it holds that $1 - \gamma \lambda_\mathbf{a} \geq 0$, and for at least one value of $\mathbf{a}$, $1 - \gamma \lambda_\mathbf{a} = 0$. 

What this implies is that, starting from the assumption of a parent with $k$ non-vanishing outcomes, we have a procedure to obtain a parent with at most $k-1$ non-vanishing outcomes. This holds whenever $k > D$, since then we are always guaranteed that the points cannot be linearly independent. We can thus iterate this procedure until we are left with a parent with $D$ outcomes (or fewer, if in the last step multiple outcomes simultaneously vanish), at which point we can no longer guarantee linear dependence between the remaining POVM elements of the parent and must terminate the procedure.

This argument is nothing but Caratheodory's theorem for cones (presented in terms of sub-measurements), which states that any point inside a convex cone in dimension $D$ can be written as a conic combination of at most $D$ extremal rays. 

\section{Appendix C: Typical complexity of the boundary of compatible measurements}
In this appendix we describe the strategy used in order to sample from the boundary of the set of compatible measurements. We induced a random measure on the boundary by using the following procedure:
\begin{enumerate}
\item Randomly generate $m$ unitary matrices $\{U_x\}$ according to the Haar measure, and from them define $m$ ideal von-Neumann measurements $\{\{\Pi_{a|x}\}_a\}_x$, via their eigenvector decomposition. These sets of measurements will be incompatible with probability one.
\item Using the dual formulation of the SDP for compatibility, extract the dual variables, which geometrically define a random direction in the space of sets of measurements
\item Find the set of compatible measurements $\{\mathbb{M}_x\}_x$ which are furthest in this direction. This problem is an SDP, and moreover will always find a set of compatible measurements on the boundary of the set, due to convexity.
\end{enumerate}
This method thus induces a measure on the boundary of compatible measurements, starting from the Haar measure on Unitary matrices. Sampling from this distribution over compatible measurements, we then estimated the probability distribution over complexity in a number of cases which were numerically tractable. The full numerical results can be found in the accompanying online notebook \cite{notebook}. 

\section{Appendix D: The correspondence between joint measurability and EPR steering}
In this appendix we provide more details on the implications of our results on related question of the complexity of Local-Hidden-State (LHS) models in the context of Einstein-Podolsky-Rosen (EPR) steering. 

EPR steering is the nonlocal effect whereby measurements performed by Alice on half of an entangled quantum state, `steer' the states of Bob at a distance in a way which cannot be explained by a simple causal model (known as LHS model). In particular, if Alice and Bob share a state $\rho^{AB}$, and Alice performs a measurement $\bM_x$, then upon obtaining outcome $a$ she steers Bob into the unnormalised state $\sigma_{a|x} = \tr_A[(M_{a|x}\otimes \openone) \rho^{AB}]$. 

The collection of states is said to have an LHS model if 
\begin{equation}
\sigma_{a|x} = \sum_\lambda p(\lambda) p(a|x,\lambda) \rho_\lambda,
\end{equation}
where $\lambda$ is a classical hidden variable, distributed according to $p(\lambda)$, $p(a|x,\lambda)$ are a collection of probabilities describing Alice's measurement outcome, and $\rho_\lambda$ are a collection of hidden normalised states, describing Bob's system. 

Similarly to the case of compatibility, one can study the complexity of LHS models. First, a `canonical' form of LHS model can always be found, whereby
\begin{equation}\label{e:LHS}
\sigma_{a|x} = \sum_{\mathbf{a}}p(\mathbf{a} )D_\mathbf{a}(a|x)\rho_{\mathbf{a}}
\end{equation}
where now the hidden variable is the tuple $\mathbf{a} = (a_1,\cdots,a_m)$, which correspond to a list of measurement results, one for each measurement of Alice, and $\rho_\mathbf{a}$ are the associated hidden states for Bob, which are jointly distributed according to $p(\mathbf{a})$. In this model, when Alice receives the hidden variable $\mathbf{a}$ and is asked to make the measurement $x$, she returns as measurement result $a = a_x$.

As in the case of compatible measurement, LHS models are inherently more complex than the assemblage of states they reproduce. In particular, the number of hidden states in the model is exponential in the number of measurement settings of Alice. We can thus ask, just as in the case of compatibility, whether it is always possible to find a simple LHS model, which contains only a small number of hidden states. In the following we will show that this is indeed the case, by exploiting the connection between steering and measurement incompatibility \cite{quintino2014,uola2014}. 

The goal here is to show that the the maximal number of states needed in any canonical LHS model is the same as the number of elements in a canonical parent POVM, i.e. 
\begin{equation}
d^2(m(o-1) + 1).
\end{equation}
The first step is to notice the one-to-one correspondence between incompatibility and steering that was recently found: every set of measurements leads to steering if and only if it is incompatible \cite{quintino2014,uola2014}. In fact, starting from an LHS model in the steering scenario, it is always possible to obtain a parent measurement, and vice-versa, using the following construction: 

\begin{itemize}
\item Notice that the states in a steering scenario satisfy the no-signalling constraint $\sum_a \sigma_{a|x} = \rho$, where $\rho$ is the reduced density operator of Bob, which is independent of $x$.
\item  Consider the purification of Bob's state  $\ket{\psi} = \sum_i \sqrt{\lambda_i}\ket{\lambda_i}\ket{\lambda_i}$, where $\rho = \sum_i \lambda_i \ket{\lambda_i}\bra{\lambda_i}$ is the diagonal form of Bob's reduced state.
\item Notice that the following operators form a collection of POVMs
\begin{equation}\label{e:meas vs assem}
M_{a|x} = \sqrt{\rho^{-1}}(\sigma_{a|x})^\mathrm{T}\sqrt{\rho^{-1}} \,,
\end{equation}
where $\mathrm{T}$ denotes transpose in the basis $\{\ket{\lambda_i}\}$.
\item Finally notice that if Alice and Bob share the state $\ket{\psi}$ (i.e., the purifying system is given to Alice), and she performs the above POVMs, then this prepares the assemblage $\sigma_{a|x}$ for Bob, 
\begin{equation}
\tr_A[(\sqrt{\rho^{-1}}(\sigma_{a|x})^\mathrm{T}\sqrt{\rho^{-1}}\otimes \openone)\ket{\psi}\bra{\psi}] = \sigma_{a|x}.
\end{equation}
\end{itemize}

Equation \eqref{e:meas vs assem} is the key equation for the one-to-one correspondence. Assume first that the assemblage $\sigma_{a|x}$ has an LHS model of the form \eqref{e:LHS}, then the associated set of measurements have the form
\begin{align}
M_{a|x} &= \sqrt{\rho^{-1}}\left(\sum_{\mathbf{a}}p(\mathbf{a})D_\mathbf{a}(a|x)\rho_\mathbf{a}^\mathrm{T}\right)\sqrt{\rho^{-1}},\nonumber \\
&= \sum_{\mathbf{a}}D_\mathbf{a}(a|x) \left[p(\mathbf{a})\sqrt{\rho^{-1}}\rho_\mathbf{a}^\mathrm{T}\sqrt{\rho^{-1}}\right] \,.
\end{align}
Defining $C_\mathbf{a} = p(\mathbf{a})\sqrt{\rho^{-1}}\rho_\mathbf{a}^\mathrm{T}\sqrt{\rho^{-1}}$, which is positive semidefinite by construction, and sums to the identity operator, we see that they constitute a canonical parent for the measurements $M_{a|x}$. In the other direction, the calculation follows identically. In particular, from \eqref{e:meas vs assem} it follows that
\begin{equation}
\sigma_{a|x} = \sqrt{\rho}M_{a|x}^\mathrm{T}\sqrt{\rho}.
\end{equation}
Hence, if $M_{a|x}$ form a set of compatible measurements, with parent satisfying by $M_{a|x} = \sum_{\mathbf{a}} D_\mathbf{a}(a|x)C_\mathbf{a}$, then
\begin{equation}
\sigma_{a|x} = \sum_{\mathbf{a}} D_\mathbf{a}(a|x) \sqrt{\rho} C_\mathbf{a}^\mathrm{T} \sqrt{\rho}.
\end{equation}
Defining $p(\mathbf{a}) = \tr[\sqrt{\rho} C_\mathbf{a}^\mathrm{T} \sqrt{\rho}]$ and $\rho_\mathbf{a} = \sqrt{\rho} C_\mathbf{a}^\mathrm{T} \sqrt{\rho}/p(\mathbf{a})$, which are seen to correspond to a valid probability distribution, and a set of normalised quantum states, we thus recover an LHS model for the assemblage $\sigma_{a|x}$.

Thus, putting everything together, whenever we have an assemblage that has an LHS model, we can always find a set of compatible measurements that reproduce the assemblage. The correspondence furthermore shows that the LHS model and the parent measurement are directly related to each other. Hence, our construction for finding a parent measurement with at most $d^2(o(m-1) + 1)$ measurements directly implies the existence of an LHS model with at most this many hidden states. 

\end{appendix}

\end{document}